\documentstyle[multicol,aps,pre,graphicx,amssymb]{revtex}
\begin{document}

\title{On the connection between financial processes with stochastic volatility 
and nonextensive statistical mechanics}
\author{S. M. Duarte Queir\'{o}s$^{1}$ and C. Tsallis$^{1,2}$ \thanks{%
E-mail addresses: {\tt sdqueiro@cbpf.br, tsallis@santafe.edu}}}
\address{$^{1}$Centro Brasileiro de Pesquisas F\'{i}sicas, Rua Dr. Xavier Sigaud 150,%
\\
22290-180, Rio de Janeiro-RJ, Brazil\\
$^{2}\,$Santa Fe Institute, 1399 Hyde Park Road,%
\\
Santa Fe,New Mexico 87501, USA}
\date{\today}
\maketitle

\begin{abstract}
The $GARCH$ algorithm is the most renowned generalisation of Engle's original proposal for modelising {\it returns}, the $ARCH$ process.
Both cases are characterised by presenting a time dependent and correlated variance or {\it volatility}. 
Besides a memory parameter, $b$, (present in $ARCH$) and an independent and identically distributed noise, $\omega $, 
$GARCH$ involves another parameter, $c$, such that, for $c=0$, the standard 
$ARCH$ process is reproduced. In this manuscript we use a generalised noise following a distribution characterised by 
an index $q_{n}$, such that $q_{n}=1$ recovers the Gaussian distribution. Matching low statistical moments of $GARCH$ 
distribution for returns with a $q$-Gaussian 
distribution obtained through maximising the entropy $S_{q}=\frac{1-\sum_{i}p_{i}^{q}}{q-1}$, basis of nonextensive 
statistical mechanics, we obtain a sole analytical connection between 
$q$ and $\left( b,c,q_{n}\right) $ which turns out to be remarkably good when compared with computational simulations. 
With this result we also derive an analytical approximation for the stationary distribution for the (squared) volatility. 
Using a generalised Kullback-Leibler relative entropy form based on $S_{q}$, we also analyse the degree of dependence 
between successive returns, $z_{t}$ and $z_{t+1}$,
of $GARCH(1,1)$ processes. This degree of dependence is quantified by an entropic index, $q^{op}$. Our analysis points the 
existence of a unique relation between the three entropic indexes $q^{op}$, $q$ and $q_{n}$ of the problem,  
independent of the value of $(b,c)$.

\end{abstract}


\begin{multicols}{2}

\section{Introduction}

\label{intro}The study of time series plays an important role in science due to their
ubiquity in both natural and artificial systems. They can be found, e.g., in geoseismic (earthquakes),
meteorological (El Ni\~{n}o), physiological (electroencephalographic
profiles) or financial phenomena \cite{abe,ausloos,plastino,osorio-borland-tsallis,apfa}. They comprise a sequentially 
ordered set of random variables, correlated or not, following a certain probability function. For the uncorrelated case,
 the most perceptive is to consider the time series as a succession of values that are associated to the same probability 
distribution, like it occurs for the ordinary random walk, where probability of a certain jump value is constant in time. 
This kind of process is defined as {\it homoskedastic}. However, there are phenomena for which the probability 
distribution associated to the stochastic variable at some time step $t$ depends explicitly on $t$ and these are named 
{\it heteroskedastic}. A simple way to obtain a heteroskedastic process is to consider the same probability functional 
for all times, but with a varying second-order moment (or width). In this sort of stochasticity we can include financial time series, 
namely return time series, where second-order moment time dependence is a feature more than well-known \cite{bouchaud-stanley}. Aiming to mimic this type of systems, Engle introduced in $1982$ the autoregressive conditional heteroskedasticity ($ARCH( s)$) ($s$ will be defined later on)  process \cite{engle}, 
which is considered a landmark in finance, comparable to the Black-Scholes equation \cite{black-scholes}, because of
its wide use \cite{boller,granger}. 
Albeit its extraordinary importance, Engle's model presents, in
some situations, implementation difficulties and above that, is not able to fully satisfactorily reproduce the 
empirically observed power-law like correlation decay of volatility. These failures lead T. Bollerslev to
generalise it defining the $GARCH(s,r) $ \cite{bollerslev} ($G$
stands for {\it generalised}). Due to its financial cradle these models are not well-known in physics, nevertheless
they can be very useful to illustrate many traditional physical problems
(see, e.g., \cite{baldovin}). In this manuscript 
we give sequence to the ansatz presented by us in a previous work \cite{arch} for the $ARCH\left(1\right) $ 
process, generalising it to the $GARCH( 1,1) $. 
Moreover, we detail the physical justification for why these models (which present time dependent variance), 
accommodate well within the current nonextensive statistical mechanical theory. We also present an analytical 
form for the distribution for the (squared) volatility and analyse the degree of dependence 
between successive {\it returns}. The manuscript is organised as follows: In Sec.~\ref{garch-model}
we introduce the $GARCH( s,r) $ process and some of its properties; In Sec.~\ref{connection}
we present our connection between $GARCH( 1,1) $ and nonextensive entropy. In addition, along the lines of 
superstatistics \cite{beck-cohen}, we derive the distribution for the (squared) volatility in Sec.~\ref{super}. In Sec.~\ref{kl}, 
applying a generalised Kullback-Leibler relative entropy we analyse the degree of dependence
between $z_{t}$ and $z_{t+1}$ elements of a $GARCH( 1,1) $ time series and its relation with non-Gaussianity. 
Some final comments are made in Sec.~\ref{remarks}.

\section{The GARCH model}

\label{garch-model}As settled by Engle \cite{engle}, we will define an
autoregressive conditional heteroskedastic time series as a discrete time
stochastic process, $z_{t}$, 
\begin{equation}
z_{t}=\sigma _{t}\,\omega _{t},  \label{z}
\end{equation}
where $\omega _{t}$ is an independent and identically distributed random
variable with null mean and unitary variance, i.e., $\left\langle \omega
_{t}\right\rangle =0$ and $\left\langle \omega _{t}^{2}\right\rangle =1$.
The quantity $\sigma _{t}$, named {\it volatility}  is time varying, positive 
defined and dependent of the past values of the {\it return} $z_{t}$.  According to its
definition, the process presents mean zero, is uncorrelated 
($\left\langle z_{t}z_{t^{\prime }}\right\rangle \sim \delta _{tt^{\prime }}$)
and has a conditional variance, $\sigma _{t}^{2}$, that evolves in time.

In his original paper Engle \cite{engle} suggests a possible expression for $\sigma
_{t}^{2}$ defining it as a linear function of past squared values of $z_{t}$
 known as $ARCH( s) $ {\it linear process},
\begin{equation}
\sigma _{t}^{2}=a+\sum_{i=1}^{s}b_{i}\,z_{t-i}^{2},\qquad \left(
a,b_{i}\geqslant 0\right) .  \label{arch}
\end{equation}
With the aim of solving the weak points of the $ARCH$ process that were referred in Sec.~\ref{intro}, it was introduced 
the linear $GARCH( s,r) $, which presents a more flexible structure for the functional form 
of $\sigma_{t}^{2}$ decreasing the direct influence of $z$ on $\sigma _{t}^{2}$ 
(for details see  \cite{bollerslev}):
\begin{equation}
\sigma
_{t}^{2}=a+\sum_{i=1}^{s}b_{i}\,z_{t-i}^{2}+\sum_{i=1}^{r}c_{i}\,\sigma
_{t-i}^{2},\qquad \left( a,b_{i},c_{i}\geqslant 0\right) ,  \label{garch}
\end{equation}
Like its predecessor, the $GARCH( s,r) $ model also captures the recognised tendency 
for {\it volatility clustering} (evident in financial time series) and is very similar to 
intermittent fluctuations in turbulent flows \cite{turbulence}: large(small) values of $z_{t}$
are usually followed by large(small) values. However,  due to the {\it arbitrary sign} of $\omega _{t}$, it can be verified that, although $\left\langle z_{t}z_{t^{\prime}}\right\rangle \sim \delta _{tt^{\prime }}$,  $\left\langle \left| z_{t}\right| \ \left| z_{t^{\prime }}\right|\right\rangle $ 
is {\it not} proportional to $\delta _{tt^{\prime }}$. For $c_{i}=0 \,(\forall i)$, $GARCH( s,r) $ straightforwardly 
reduces to the linear $ARCH(s) $. See Figs. 1, 2 and 3 for typical realisations.
\begin{figure}[tbp]
\begin{center}
\includegraphics[width=0.75\columnwidth,angle=0]{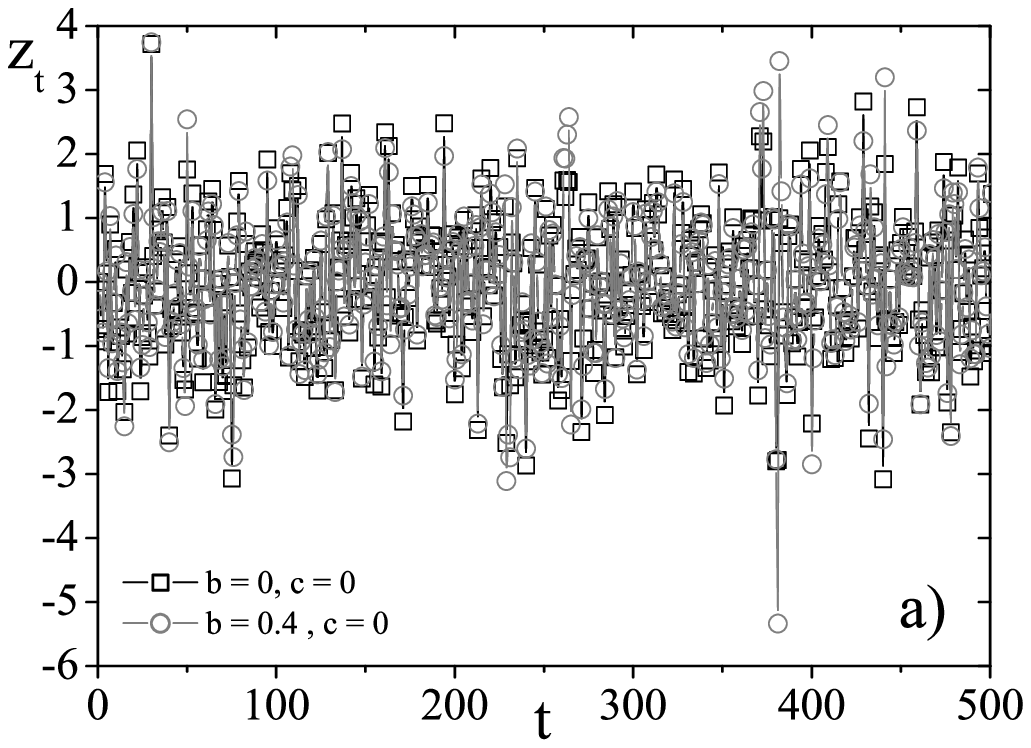}
\includegraphics[width=0.75\columnwidth,angle=0]{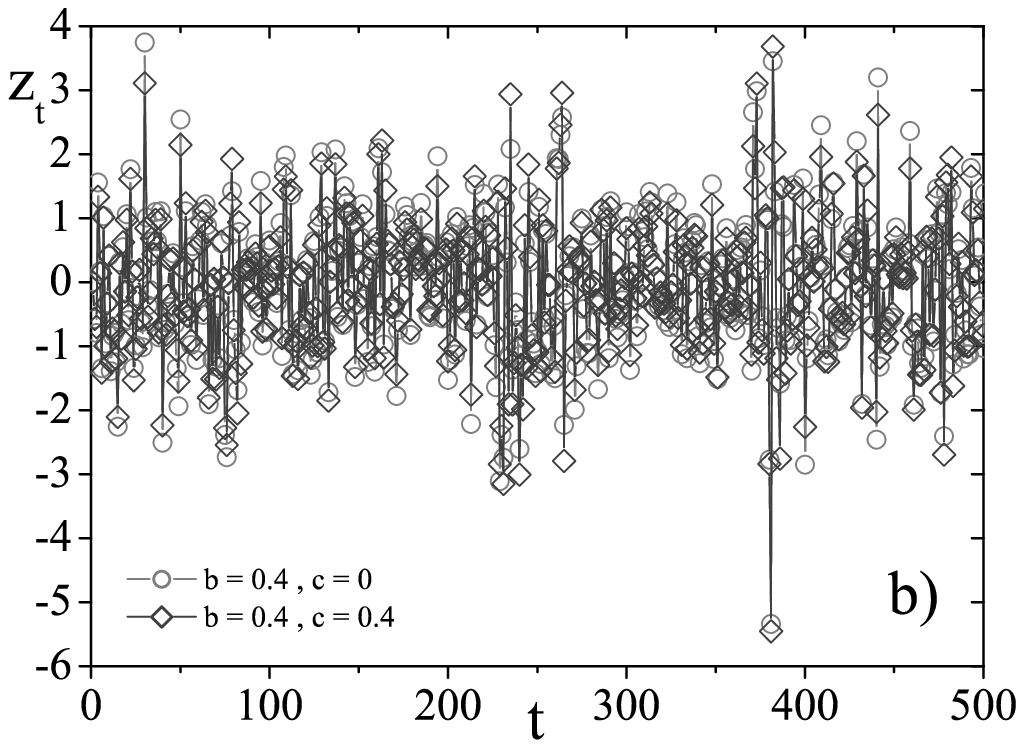}
\end{center}
\caption{Typical $GARCH\left( 1,1\right) $ time series obtained for a
Gaussian noise, $q_{n}=1$. In (a) we present  time series which correspond
to pure Gaussian ($\Box $) and $ARCH\left( 1\right) $ ($\circ $).
In (b) the introduction of parameter $c$  ($\diamond $) increases the
probablility for larger values of $\left| z_{t}\right| $, thus increasing tails in 
$P\left( z\right) $.}
\label{fig-1}
\end{figure}
\begin{figure}[tbp]
\begin{center}
\includegraphics[width=0.75\columnwidth,angle=0]{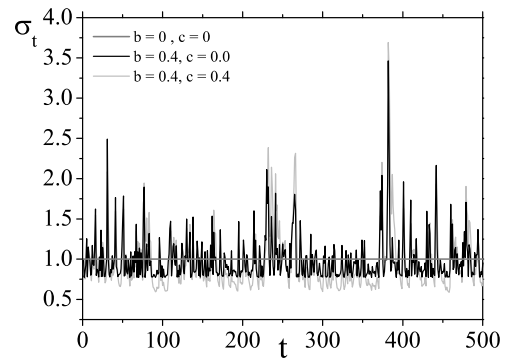}
\end{center}
\caption{Time dependence of volatility $\protect\sigma $ for the $GARCH$
processes presented in Fig.~\ref{fig-1}. \ Here is clearly visible the
difference between $c=0$ ($ARCH$) and $c\neq 0$ ($GARCH$). For the same values of $b$ we are able to obtain 
greater values of $\protect\sigma $, thus leading to fatter tails in $P\left( z\right) $.}
\label{fig-2}
\end{figure}
\begin{figure}[tbp]
\begin{center}
\includegraphics[width=0.75\columnwidth,angle=0]{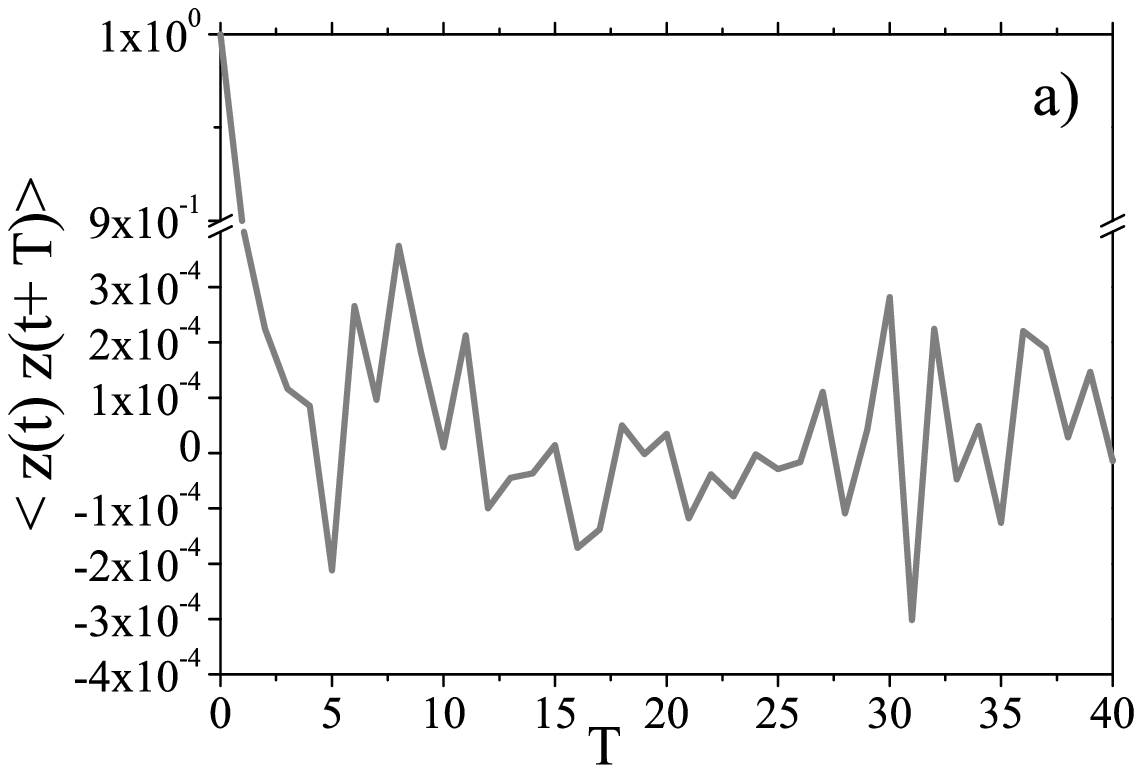}
\includegraphics[width=0.75\columnwidth,angle=0]{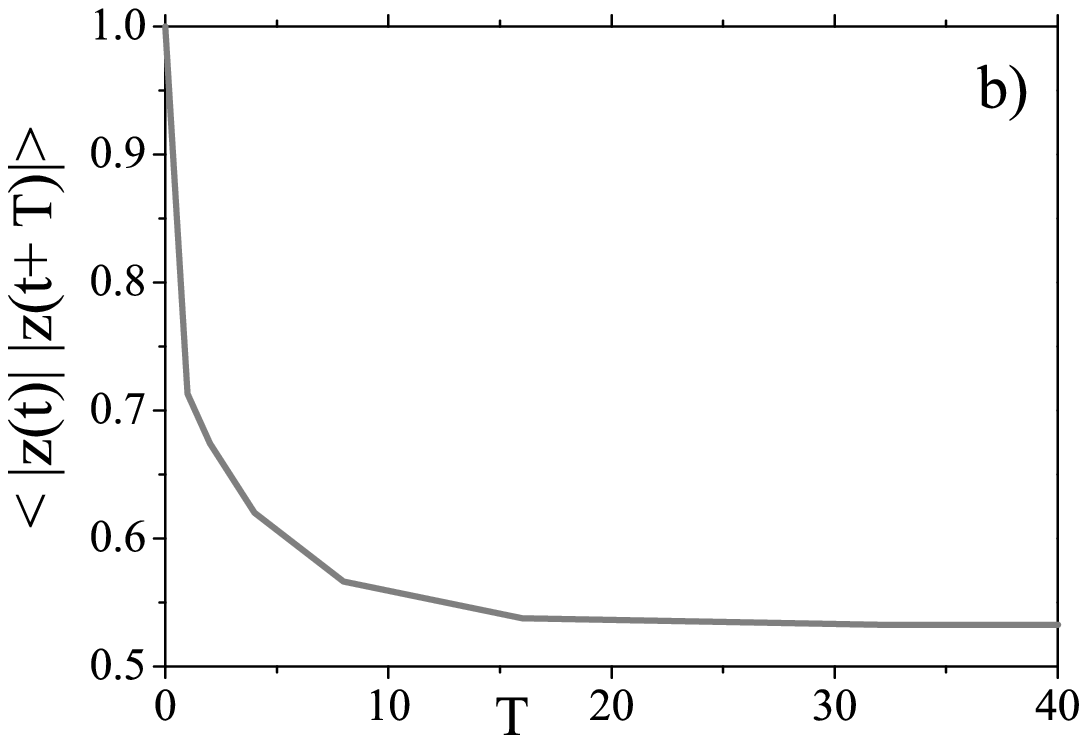}
\end{center}
\caption{In (a) correlation between returns $z(t+T)$ and $z(t)$ vs. time lag $%
T$ . All values, except for $T=0$, are at noise level. \ In (b) correlation between absolute
values of returns which present a decay of time lag $T$. The run analysed
was the same presented in Fig.~\ref{fig-1}}
\label{fig-3}
\end{figure}
Let us focus on the simplest and most used of the $GARCH$
processes, namely the $GARCH( 1,1) $ one. In this case, 
\begin{equation}
\sigma _{t}^{2}=a+b\,z_{t-1}^{2}+c\,\sigma _{t-1}^{2},  \label{garch11}
\end{equation}
so that the process is completely defined when $a$, $b$, $c$,
and the noise nature, $P_{n}\left( \omega _{t}\right) $, are specified.

Combining Eqs.~(\ref{z}) and (\ref{garch11}) we obtain the $n^{th}$ order moment for
the stationary $P\left( z\right) $ distribution, particularly the second,
\begin{equation}
\bar{\sigma}^{2}\equiv \left\langle z_{t}^{2}\right\rangle =\left\langle
\sigma _{t}^{2}\right\rangle =\frac{a}{1-b-c},\qquad \left( b+c<1\right) ,
\label{2m}
\end{equation}
and the fourth moment, 
\begin{equation}
\left\langle z_{t}^{4}\right\rangle =a^{2}\,\left\langle \omega
_{t}^{4}\right\rangle \frac{1+b+c}{\left( 1-b-c\right) \left(
1-2\,b\,c-c^{2}-b^{2}\,\left\langle \omega _{t}^{4}\right\rangle \right) }.
\label{4m}
\end{equation}
Let us now assume, for simplicity and without lack of generality, a 
$GARCH\left( 1,1\right) $ process that generates time series with unitary
variance, i.e., $\bar{\sigma}^{2}=1$, which imposes $a=1-b-c$. Now,
for this process, the fourth moment is numerically equal to the kurtosis 
$k_{x}\equiv \frac{\left\langle x^{4}\right\rangle }{\left\langle x^{2}\right\rangle ^{2}}$, 
and thus we get, 
\begin{equation}
\left\langle z_{t}^{4}\right\rangle =k_{z}=k_{\omega }\left( 1+b^{2}\frac{%
k_{w}-1}{1-c^{2}-2\,b\,c-b^{2}\,k_{\omega }}\right) ,  \label{kz}
\end{equation}
where $c^{2}+2\,b\,c+b^{2}\,k_{\omega }<1$.
It is clear from Eq.~(\ref{kz}) that $GARCH( 1,1) $ generates
distributions $P\left( z\right) $ with fatter tails than those of the noise $\omega _{t}$,
and fatter also than the ones obtained with $ARCH( 1) $ with the same $b$ \cite{embrechts}. 
It is also possible to see that parameter $c$ is only useful for $b\neq 0$,
otherwise $GARCH( 1,1) $ process reduces to constructing stationary
probability distributions with a kurtosis $k_{z}$ equal to $k_{w}$.

\section{The ansatz connecting GARCH and nonextensive statistical mechanics}

\subsection{Stationary distribution for returns}

\label{connection}We shall now establish a connection --- physically motivated  
in the next subsection --- between the present stochastic process and the 
current nonextensive statistical mechanical framework, based on the entropic form \cite{tsallis}, 
\begin{equation}
S_{q}=\frac{1-\int\nolimits_{-\infty }^{+\infty }\left[ p(z)\right]
^{q}dz}{q-1},\qquad \left( q\in \Re \right) .  \label{sq}
\end{equation}
This entropy is currently referred to as {\it nonextensive} because it is so for {\it independent} subsystems. It can 
however be {\it extensive} in the presence of scale-invariant correlations \cite{nota,tsallis-erice,sato-tsallis,tsallisgellmannsato}. The 
associated statistics has been successfully applied to phenomena presenting some 
kind of scale-invariant geometry, like in low-dimensional dissipative and conservative 
maps \cite{mapas}, anomalous (correlated) diffusion \cite{fokker}, turbulent flows 
\cite{beck-lewis-swinney}, Langevin dynamics with fluctuating temperature \cite{beck-cohen,wilk,beck}, 
long-range many-body classical Hamiltonians \cite{rotores}, among many others\cite{gm-ct}. Entropy (\ref{sq}) constitutes a 
generalisation of the Boltzmann-Gibbs (BG) one, namely
$ S_{BG}=-\int\nolimits_{-\infty }^{+\infty }p(z)\,\ln p(z)\,dz $. Indeed, this celebrated expression is recovered as the  
$q\rightarrow 1$ limit of entropy (8).

By applying the standard variational principle on entropy (8) with the constraints $\int\nolimits_{-\infty }^{+\infty }p(z)\,dz=1$ 
and 
$\int\nolimits_{-\infty }^{+\infty }z^{2}\,[p(z)]^q \,dz / \int\nolimits_{-\infty }^{+\infty }\,[p(z)]^q \,dz =\bar{\sigma}^{2}_{q} $ 
\cite{tsallis-renio-plastino,pratotsallis,tsallis-bjp} 
($\bar{\sigma}^{2}_{q}$ is defined as the {\it generalised second-order moment}) we obtain
\begin{equation}
p(z)=\frac{{\cal A}}{\left[ 1+{\cal B\,}\left( q-1\right) \,z^{2}%
\right] ^{\frac{1}{q-1}}},\qquad \left( q<\frac{5}{3}\right) ,  \label{plz}
\end{equation}
where
\begin{equation}
{\cal B} \equiv \frac{1}{\bar{\sigma}^{2}\,\left( 5-3\,q\right) },  \label{b}
\end{equation}
\begin{equation}
\bar{\sigma}^{2} \equiv \int\nolimits_{-\infty }^{+\infty }p(z)\,dz ,  \label{2momento}
\end{equation}
and 
\begin{equation}
{\cal A}=\frac{\Gamma \left[ \frac{1}{q-1}\right] }{\sqrt{\pi }\,\Gamma %
\left[ \frac{3-q}{2q-2}\right] }\sqrt{\left( q-1\right) \,{\cal B}}.
\label{a}
\end{equation}
Standard and generalised second-order moments, $\bar{\sigma}^{2}$ and $\bar{\sigma}^{2}_{q}$, are related by 
$\bar{\sigma}^{2} = \bar{\sigma}^{2}_{q} \frac{3-q}{5-3\,q}$.
In the limit $q \to 1$, distribution (9) becomes a Gaussian. If $q=\frac{3+m}{1+m} \,(m=1,2,3...)$, distribution (9) recovers the 
Student's t-distribution with $m$ degrees of freedom; if $q=\frac{n-4}{n-2}  \, (n=3,4,5...)$, 
it recovers the so-called $r$-distribution with $n$ degrees of freedom \cite{souzatsallis}.  

\begin{figure}[tbp]
\begin{center}
\includegraphics[width=0.75\columnwidth,angle=0]{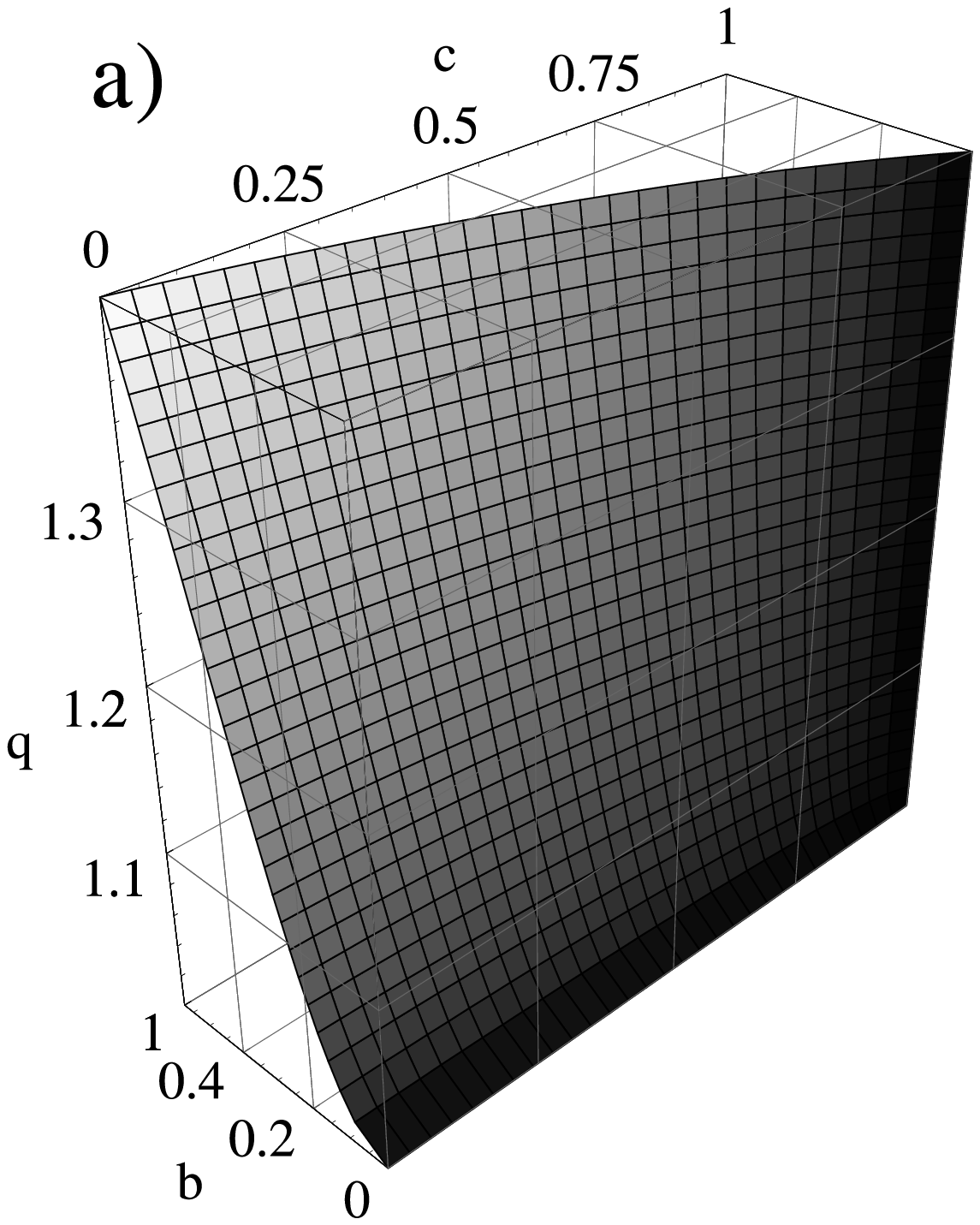}
\includegraphics[width=0.75\columnwidth,angle=0]{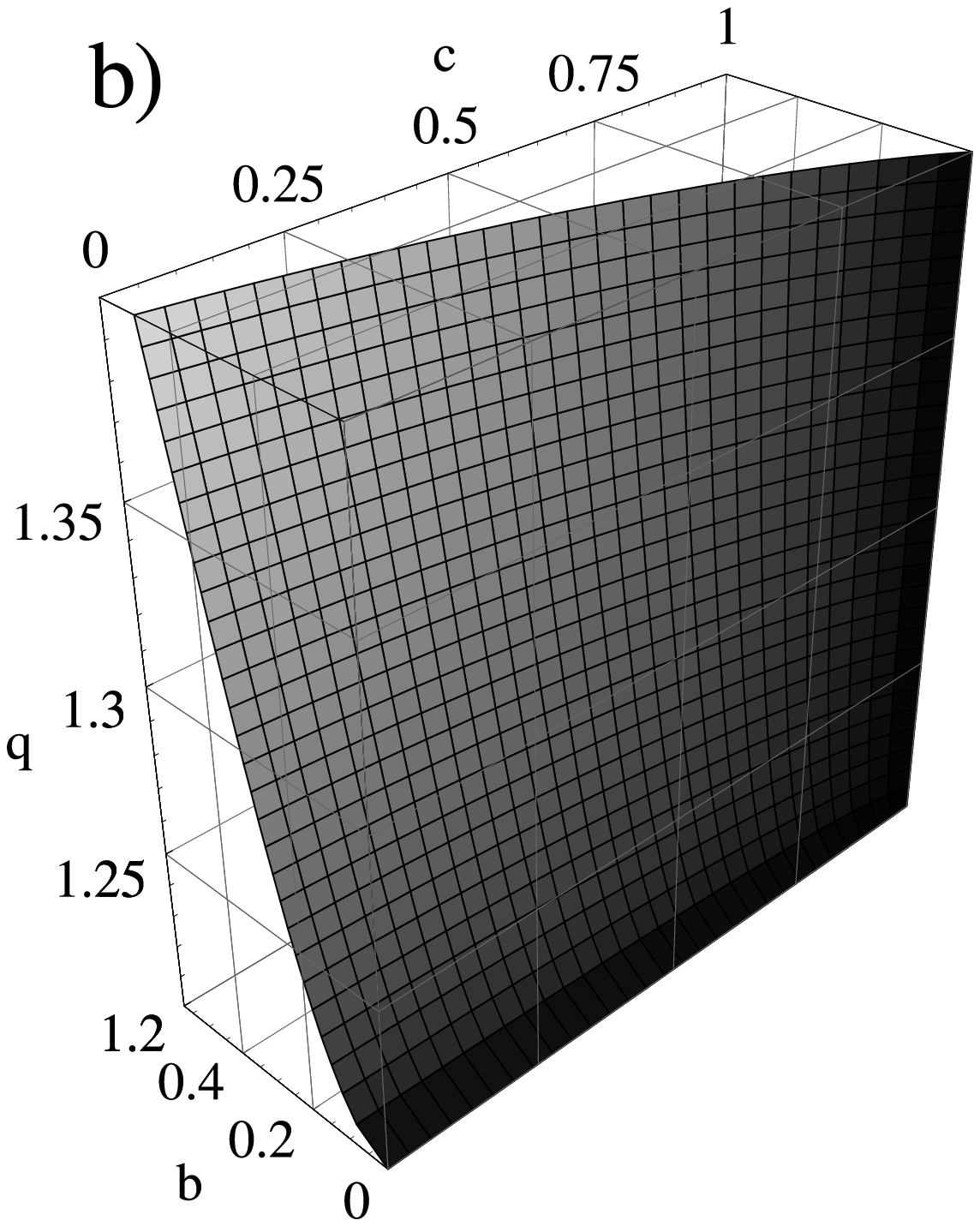}
\end{center}
\caption{Diagram $(q,b,c)$ for $q_{n}=1$ in a) and $q_{n}=1.2$ in b). In b) we can see that the greatest allowed value of $b$ ($c=0$) 
is $b = \frac{1}{\sqrt{4.2}} \simeq 0.488$.
}
\label{fig-4}
\end{figure}

Defining the $q${\it -exponential} function as,
\begin{equation}
exp_{q}(x)\equiv [1+(1-q)x]^{\frac {1}{1-q}}  \;\;\;\;(exp_1(x)=e^x)             \,, \label{q-expf}
\end{equation}
we can rewrite the above distribution as follows,
\begin{equation}
p(z)= {\cal A}\,e_{q}^{-\,{\cal B}\,z^{2}} \,,
\label{pz}
\end{equation}
from now on referred to as $q${\it -Gaussian}. The fourth moment of $p(z)$ is,
\begin{equation}
\left\langle z^{4}\right\rangle =3\,\left( \bar{\sigma}^{2}\right) ^{2}%
\frac{3\,q-5}{5\,q-7}.  \label{kzq}
\end{equation}
Let us now propose the ansatz $p(z)\simeq P(z)$ with $ \bar{\sigma}^{2}=1$. 
Specifically, we will impose the matching of
Eqs.~(\ref{kz}) and (\ref{kzq}). We assume that the noise $\omega _{t}$ follows the
generalised distribution
\begin{equation}
P_{n}(\omega )=\frac{{\cal A}_{q_{n}}}{\left[ 1+ \, \frac{q_{n}-1}{5-3q_n} \, \omega ^{2}%
\right] ^{\frac{1}{q_{n}-1}}},\qquad \left( q_{n}<\frac{5}{3}\right) ,  \label{pnoise}
\end{equation}
defined by the index $q_{n}$; its variance equals unity, and $ {\cal A}_{q_{n}}$ 
is uniquely determined through normalization. We are then able to establish a 
relation between parameters $b$, $c$ and indices $q_{n}$ and $q$: 
\begin{equation}
\begin{array}{c}
b=\frac{\sqrt{( q-q_n) [            ( 5- 3q_n) (2- q)     -c^{2}( 5 -3q) (2- q_n)         ] }}{(5-3q_n) (2- q)  } 
\\ 
\\ 
  -\frac{c \, ( q-q_n) }{(5- 3q_n) ( 2-q) } \,.
\end{array}
 \label{ansatz}
\end{equation}
For $b=c=0$ we have $q=q_n$. 
Naturally, the connection indicated in Eq. (17) and illustrated in Fig.~\ref{fig-4} for typical values of $q_{n}$  reduces, for $c=0$, 
to the one obtained \cite{arch} for
the $ARCH\left( 1\right) $ model, namely $b=\sqrt{(q-q_n)/[(5-3q_n)(2-q)]}$, hence $q=[q_n+2b^2(5-3q_n)]/[1+b^2(5-3q_n)]$.

\begin{figure}[tbp]
\begin{center}
\includegraphics[width=0.75\columnwidth,angle=0]{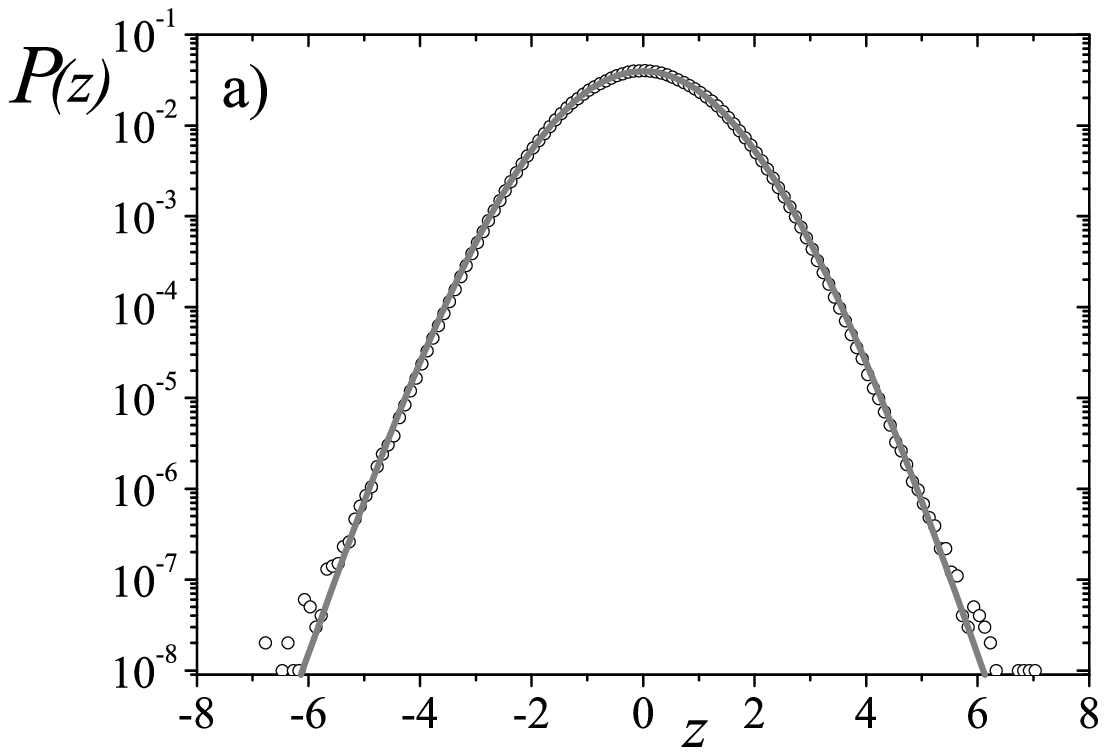}
\includegraphics[width=0.75\columnwidth,angle=0]{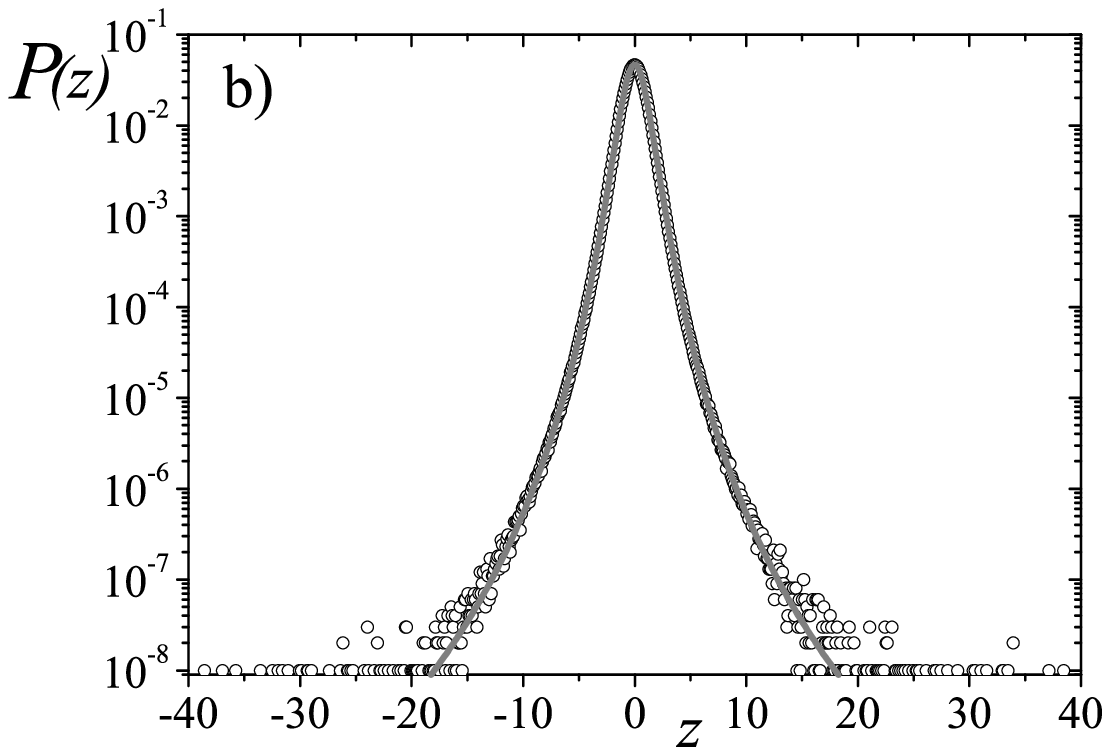}
\includegraphics[width=0.75\columnwidth,angle=0]{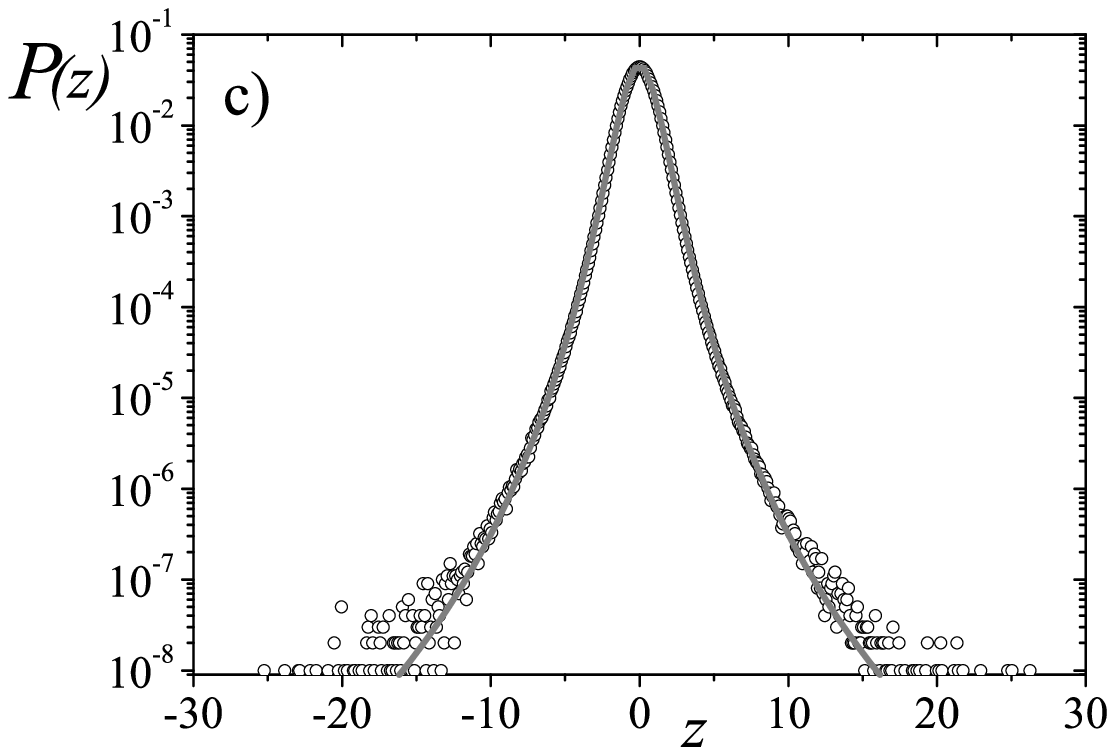}
\includegraphics[width=0.75\columnwidth,angle=0]{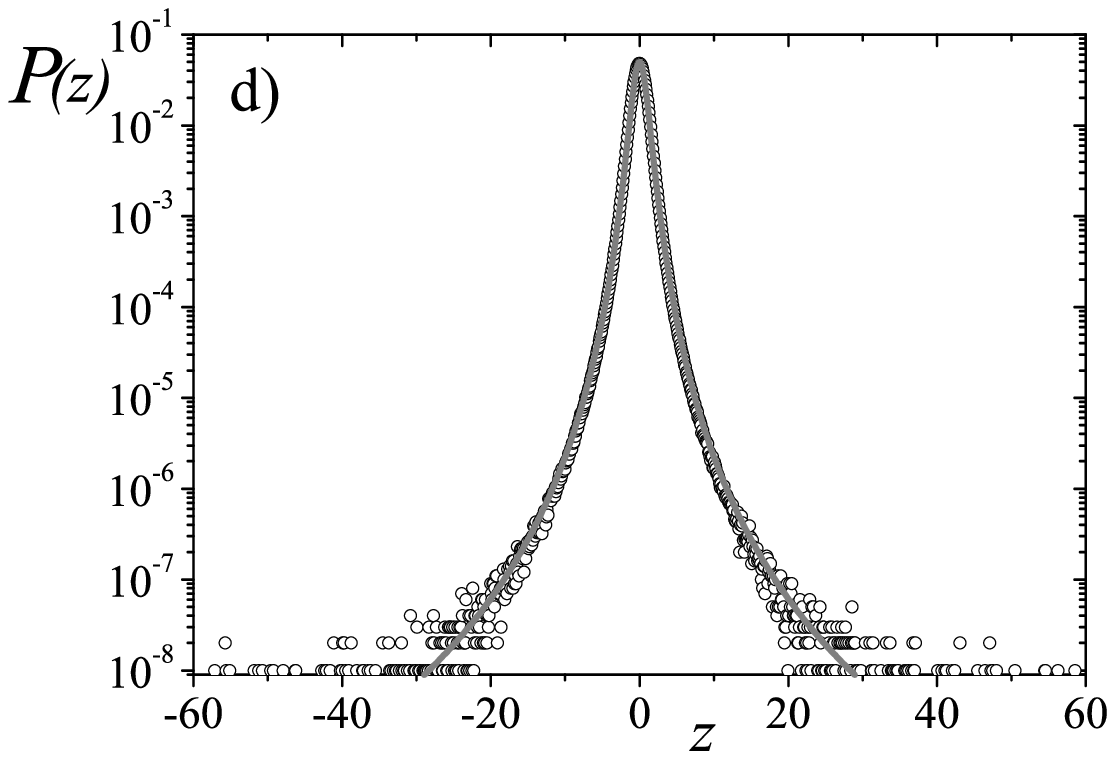}
\end{center}
\caption{PDFs for a $q_{n}=1$ noise and typical values of $\left(
b;\,c\right) $ pair. a) $\left( 0.1;\,0.1\right) $, $q=1.021$ ($\protect\chi %
^{2}=2.35\times 10^{-9}$); b) $\left( 0.1;\,0.88\right) $, $q=1.287$ ($%
\protect\chi ^{2}=4.59\times 10^{-10}$); c) $\left( 0.4;\,0.1\right) $, $%
q=1.26$ ($\protect\chi ^{2}=2.44\times 10^{-9}$); d) $\left(
0.4;\,0.4\right) $, $q=1.38$ ($\protect\chi ^{2}=3.22\times 10^{-7}$). }
\label{fig-5}
\end{figure}

\begin{figure}[tbp]
\begin{center}
\includegraphics[width=0.75\columnwidth,angle=0]{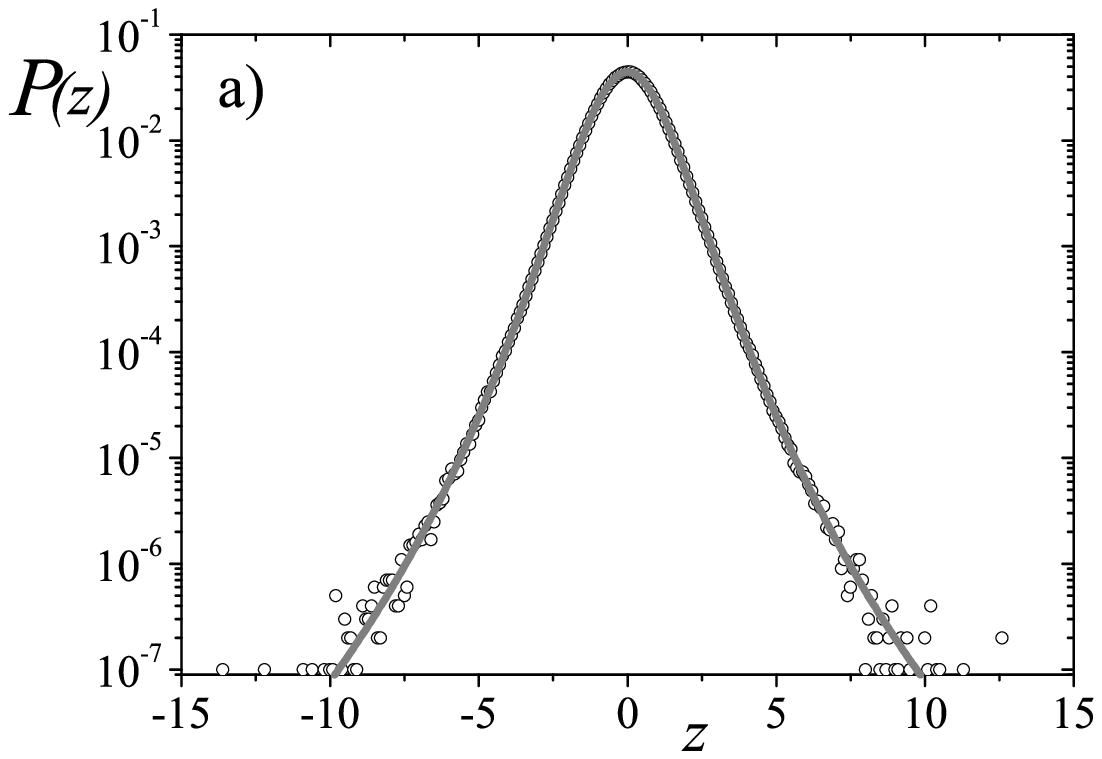}
\includegraphics[width=0.75\columnwidth,angle=0]{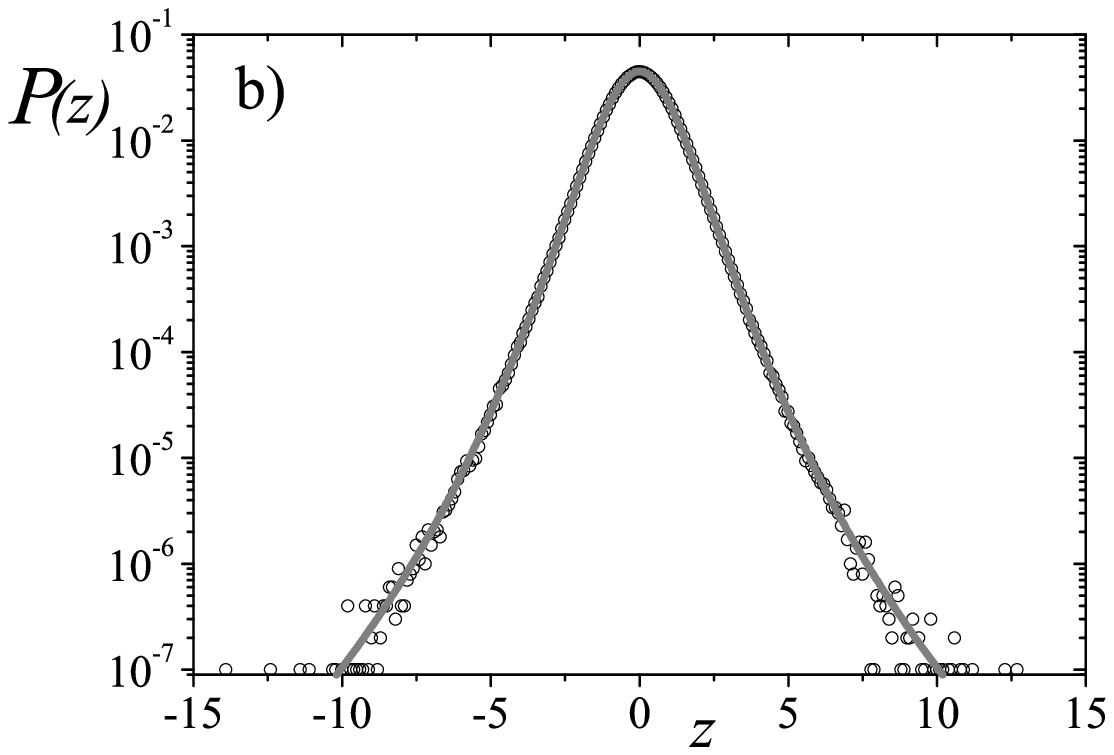}
\includegraphics[width=0.75\columnwidth,angle=0]{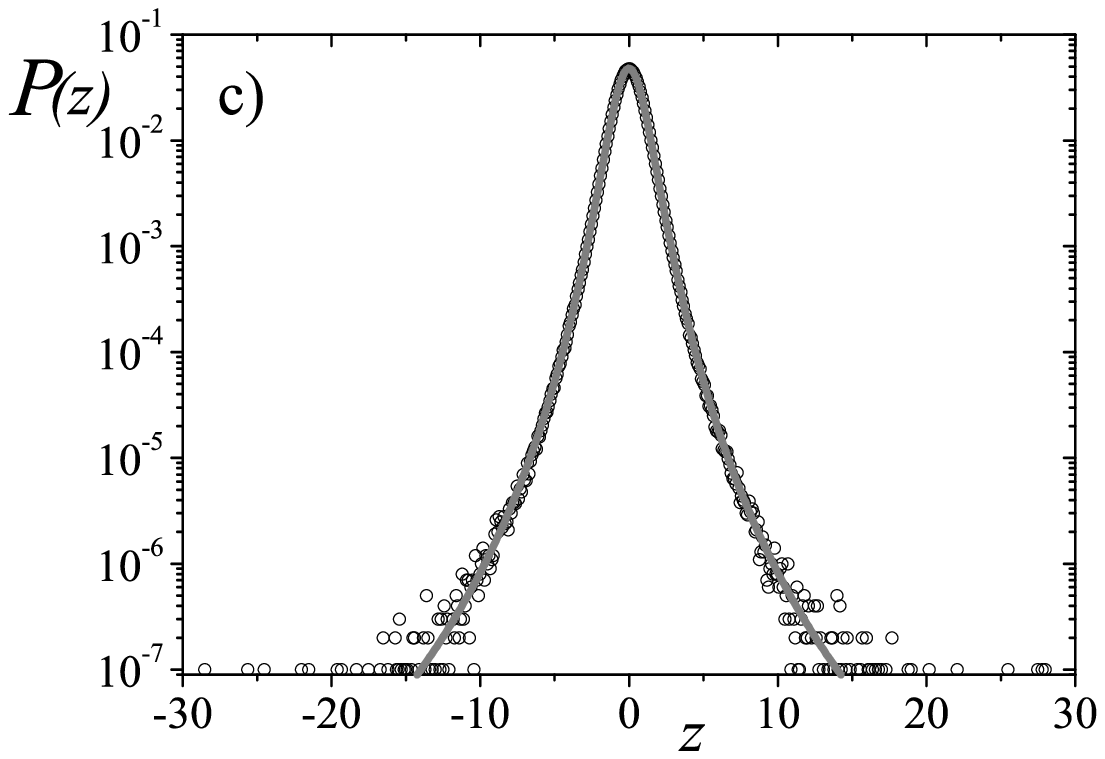}
\includegraphics[width=0.75\columnwidth,angle=0]{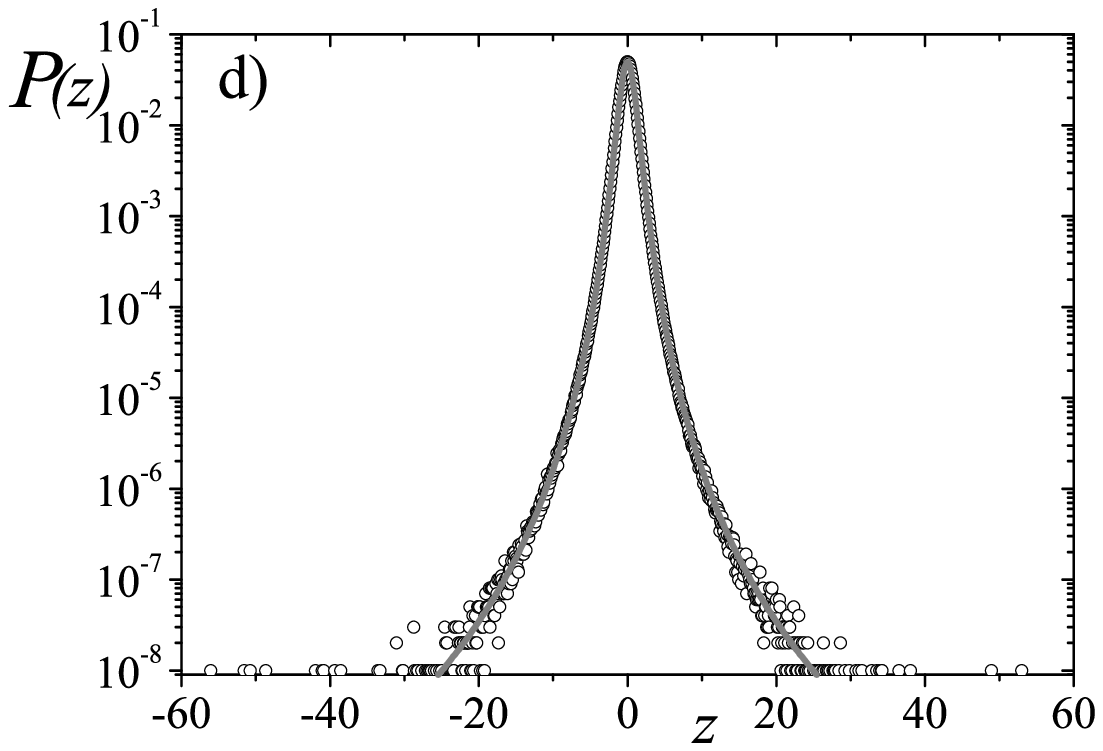}
\end{center}
\caption{PDFs for a $q_{n}=1.2$ noise and typical values of $\left(
b;\,c\right) $ pair. a) $\left( 0.1;\,0.1\right) $, $q=1.211$ ($\protect\chi %
^{2}=8.67\times 10^{-10}$); b) $\left( 0.1;\,0.5\right) $, $q=1.221$ ($%
\protect\chi ^{2}=6.01\times 10^{-10}$); c) $\left( 0.3;\,0.25\right) $, $%
q=1.310$ ($\protect\chi ^{2}=8.11\times 10^{-9}$); d) $\left(
0.3;\,0.45\right) $, $q=1.35$ ($\protect\chi ^{2}=7.36\times 10^{-9}$).}
\label{fig-6}
\end{figure}

 To verify the above ansatz we generated, for typical values of $(b,c,q_{n})$ and using an algorithm
based on Eqs.~(\ref{z}) and (\ref{garch11}), a set of $GARCH$ time series. Then we computed
the corresponding probability density functions and compared them with
the histograms (with any adequately chosen interval $\delta $) associated with the
$q$-Gaussian distribution with $q$ satisfying the ansatz.
We compared then the numerical probability density functions (PDFs) with $H( z) $ 
\begin{equation}
\begin{array}{c}
H( z) =  \int\nolimits_{z\,+\,\delta /2}^{z\,+\,\delta /2}p\left( x\right) \,dx=\frac{%
\Gamma \left[ \frac{1}{q-1}\right] }{2\,\Gamma \left[ \frac{1}{q-1}-\frac{1}{%
2}\right] }\sqrt{\frac{1-q}{\pi \,\left( \,3\,q-5\right) }} \\ 
\\ 
\times \left\{ \left( \delta -2\,z\right) \,_{2}F_{1}\left( \frac{1}{2},%
\frac{1}{q-1};\frac{3}{2};\frac{\left( q-1\right) \,\left( \delta
-2\,z\right) ^{2}}{4\,\left( 3\,q-5\right) }\right) \right.  \\ 
\\ 
\left. +\left( \delta +2\,z\right) \,_{2}F_{1}\left( \frac{1}{2},\frac{1}{q-1%
};\frac{3}{2};\frac{\left( q-1\right) \,\left( \delta +2\,z\right) ^{2}}{%
4\,\left( 3\,q-5\right) }\right) \right\} , 
\end{array}
 \label{pdf}
\end{equation}
where $_{2}F_{1}$ is the hypergeometric function.
As can be seen in Figs.~\ref{fig-5} and \ref{fig-6}, the accordance between
stationary PDFs and the PDFs obtained by 
using, in Eq.~(\ref{plz}), the value of $q$ satisfying Eq.~(\ref{ansatz}),  is quite satisfactory. 
In the captions of Figs.~\ref{fig-5} and \ref{fig-6}, we present the values of the $\chi ^{2}$ error function, 
$$\chi ^{2} \equiv \frac{1}{N}\sum\limits_{i=1}^{N}\left[ P\left( z\right) -H\left( z\right) \right] ^{2}. $$ 
Another way to evaluate the slight discrepancy between $P\left( z\right) $ and $p\left( z\right) $ (or $H(z)$) is
to compute the percentual error in the sixth-order moment between PDFs. The
results presented in Tab.~\ref{tab-1} show that discrepancies are never larger than $3\%$, 
which in practice can be considered negligible. It is interesting to notice
the remarkable agreement at least down to $p\left( z\right) =10^{-6}$ (typical
limit used in finance, see Ref. \cite{osorio-borland-tsallis}).

\subsection{Stationary distribution for (squared) volatility}

\label{super}
\begin{figure}[tbp]
\begin{center}
\includegraphics[width=0.75\columnwidth,angle=0]{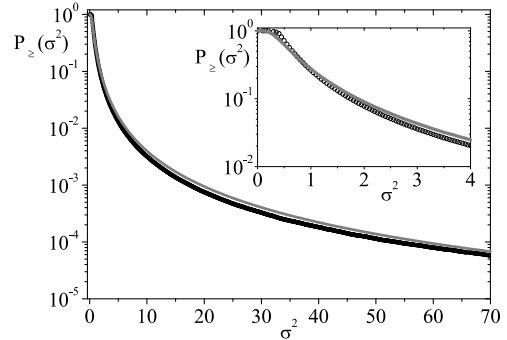}
\end{center}
\caption{The symbols in black represent the cumulative probability
distribution, $P_{\geqslant }\left( \protect\sigma ^{2}\right) $ obtained from numerical simulation
for a Gaussian noise with $b=c=0.4$ . The
gray line represents the same distribution as given by Eq. (22) with
%using Eqs.~(\ref{sigma-grande}), (\ref{sigma-medio}) and (\ref{sigma-4}). 
$(\kappa, \lambda, \bar{\sigma}^{2}) =(0.444,2.125,1)$ satisfying Eqs. (23) and (24).} 
\label{fig-7}
\end{figure}

The good results of the proposal presented in the previous sub-section have their explanation in the main feature 
of the ARCH processes, namely the temporal dependence of the variance. This fact allow us to think
of them along the same lines of Wilk and W\l odarczyk \cite{wilk} 
and of  Beck \cite{beck}. They essentially considered a system following a Langevin 
equation with fluctuating temperature, genesis of the {\it superstatistics} (statistics of statistics) recently advanced by Beck and Cohen \cite{beck-cohen}. 
This  approach was developed to treat non-equilibrium systems composed, for instance, of smaller cells obeying BG statistics with a distribution $P_{BG} \propto e^{-\beta E}$, $E$ being the energy. The long-term stationary  state 
that presents a spatio/temporaly fluctuating temperature following a distribution $f(\beta )$. 
In the long-term, the probability density for the nonequilibrium system comes from BG statistics associated 
with the small cells that are averaged over the various $\beta $, i.e.,
\begin{equation}
P_{stationary}(E)=\int f\left( \beta \right) \,P_{BG}(E) \,d\beta\;.
\end{equation}
Instead of defining the intensive fluctuating parameter, $\beta $, as the inverse temperature, 
we will define it as an inverse second-order moment, $\beta _{\sigma }=\frac{1}{2\, \sigma ^{2}}$. 
For long times, we can study  $GARCH$ as a stationary diffusion process, composed by $t$ increments, each of them of size $z_{t}$. The total value is given by $x_{t}\equiv \sum\limits_{i=1}^{t}\,z_{i}$, where each increment 
follows a certain $q_{n}-Gaussian$, with a second-order moment, 
$\sigma ^{2}$, that is associated with a probability
distribution, $p_{\sigma }( \sigma ^{2}) $. The stationary probability distribution $p( z) $ is thus given by 
\begin{equation}
p( z) =\int_{0}^{\infty }p_{\sigma }( \sigma ^{2}) \
p( z | \sigma ^{2}) \ d( \sigma ^{2}) ,
\label{superstatistics}
\end{equation}
where $ p( z | \sigma ^{2}) $ is the {\it conditional} probability of having a value $z$ for the return given a value
$ \sigma ^{2}$. The homoskedastic case corresponds to $p_{\sigma }\left( \sigma ^{2}\right) =\delta \left( \sigma 
^{2}-\bar{\sigma}^{2}\right) $.
Let us focus, for now, on the case $q_{n}=1$ (Gaussian noise) with
\begin{equation}
p( z|\sigma ^{2}) =\frac{1}{\sqrt{2\,\pi \,\sigma ^{2}}}\,
e^{-\frac{z^{2}}{2\,\sigma ^{2}}}.
\end{equation}
In their proposal Beck and Cohen \cite{beck-cohen} showed that, if the intensive parameter $\beta $ 
is associated with a Gamma distribution, then the macroscopic non-equilibrium 
steady state follows {\it exactly} a $q-${\it exponential} distribution (see also \cite{wilk,beck}). Following the reverse line and assuming 
$p(z)= P_{stationary} \simeq P(z)$, we are lead to a Gamma distribution in $\beta _{\sigma }$. In other words,
\begin{equation}
p_{\sigma }\left( \sigma ^{2}\right) =\frac{\exp \left( -\frac{1}{2\ \kappa
\ \sigma ^{2}}\right) \left( \sigma ^{2}\right) ^{-1-\lambda }}{\left( 2\
\kappa \right) ^{\lambda }\ \Gamma \left[ \lambda \right] }  \label{ps} \,,
\end{equation}
where 
\begin{equation}
\lambda = \frac{1}{q-1}-\frac{1}{2}  
\label{lambda}
\end{equation}
and
\begin{equation}
\kappa = \frac{1-q}{\bar{\sigma}^{2}\left( 3\,q-5\right) }  \,.
\label{kappa}
\end{equation}
As can be seen in Fig.~\ref{fig-7}, the ansatz gives also a quite 
satisfactory description for the probability distribution in the (squared)
volatility, corroborating the connection between the ARCH class of processes, nonextensive statistical 
mechanics, and superstatistics. The analytic expressions (\ref{ps}), (\ref{lambda}) and (\ref{kappa}) can be
very useful in applications related, among others, to option prices \cite{osorio-borland-tsallis,borland} 
where volatility forecasting plays a particularly important role \cite{willmott,borland-bouchaud}.

For the case of a $q_{n}$-Gaussian noise (with $q_{n}>1$) similar arguments can
be used. However, the achievement of an analytical solution for Eq.~(\ref{superstatistics}) which generalises 
the approach above is not a trivial task. 

For the conditional probability 
$p( z\ |\sigma ^{2}) $ we know a satisfactory answer \cite{transfvar}, namely 
\begin{equation}
p( z| \sigma ^{2}) =\frac{{\cal A}_{( q_{n}, \sigma ^{2}) }}{\left[ 1+{\cal B}_{\left( q_{n}, \sigma 
^{2}\right) }{\cal \,}\left( q_{n}-1\right) \,z^{2}\right] ^{\frac{1}{q_{n}-1}}}  \,. \label{pzq-qn}
\end{equation}

But the same does not happen with $p_{\sigma }\left( \sigma ^{2}\right) $.
Assuming a continuous approach in $q_{n}$, a good ansatz for describing 
$p_{\sigma }\left( \sigma ^{2}\right) $ is
\begin{eqnarray}
%\begin{array}{c}
p_{\sigma }\left( \sigma ^{2}\right) &\propto& \left( \,\sigma ^{2}\right) ^{-1-\lambda }\exp _{q_{\sigma }}\left( -\frac{1}{2\ \kappa \ \sigma ^{2}}\right)  \nonumber \\
&=& \left( \,\sigma ^{2}\right)
^{-1-\lambda }\left( 1+\frac{q_{\sigma }-1}{2\,\kappa }\frac{1}{\sigma ^{2}}%
\right) ^{\frac{1}{1-q_{\sigma }}} \,,
%\end{array}
 \label{ps-qn}
\end{eqnarray}
where $q_{\sigma }$ is an index which depends on $q_{n}$, i.e., $q_{\sigma }(q_{n})$ such that $q_{\sigma }(1)=1$. 
For large $\sigma ^{2}$, 
\begin{equation}
p_{\sigma }\left( \sigma ^{2}\right) \sim \left( \sigma ^{2}\right)
^{-1-\lambda } \,.  \label{sigma-grande}
\end{equation}
The integral $\int_{0}^{\infty }\sigma ^{2}\,p_{\sigma }( \sigma^{2}) \,d( \sigma ^{2}) $ should equal 
the mean value $\bar{\sigma}^{2}=\frac{a}{1-b-c}$. This yields, 
\begin{equation}
\frac{1+\lambda -\lambda \,q_{\sigma }}{2\,\kappa \,\left( \lambda -1\right) 
}=\bar{\sigma}^{2},  \label{sigma-medio}
\end{equation}
and
\begin{equation}
\int_{-\infty }^{\infty }z^{4}\int_{0}^{\infty }p_{\sigma }\left( \sigma
^{2}\right) \ p\left( z\ |\ \sigma ^{2}\right) \ d\left( \sigma ^{2}\right)
\,dz=3\,\left( \bar{\sigma}^{2}\right) ^{2}\frac{3\,q-5}{5\,q-7}.
\label{sigma-4}
\end{equation}

\begin{figure}[tbp]
\begin{center}
\includegraphics[width=0.75\columnwidth,angle=0]{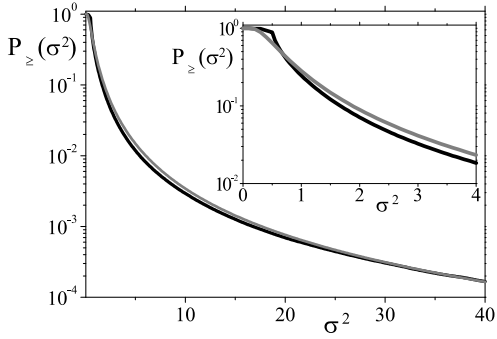}
\end{center}
\caption{The line in black represent the cumulative probability
distribution, $P_{\geqslant }\left( \protect\sigma ^{2}\right) $ obtained from
numerical simulation for a $
q_{n}-$Gaussian noise ($q_{n}=1.15$) with $(b,c)=(0.5,0)$. The gray line represents the same distribution as given by Eq. (26) with
%for the same process using Eqs.~(\ref{ps}) and (\ref{lambda-kappa}). 
$(\kappa,\lambda, \bar{\sigma}^{2})=(0.365,2.371,1)$, and $q_{\sigma }=1$.}
\label{fig-8}
\end{figure}

From Eq.~(\ref{sigma-grande}) and by adjusting numerically the curves for the cumulated
probability distributions, $P_{\geqslant }\left( \sigma ^{2}\right) $, we
were able to determine $\lambda $, and then $q_{\sigma }$ and $\kappa $ from
Eqs.~(\ref{sigma-medio}) and (\ref{sigma-4}). The procedure appears to be
valid for values of $q_{n}$ close to unity. From the analysis of some values of
$q_{n}$ we verified that the value of $q_{\sigma }$ equals $1$ for every $q_{n}$ considered.
Figures~\ref{fig-7} and \ref{fig-8} confirm that our proposal produces a satisfactory approximation when 
compared with numerical simulations, particularly for large values of 
volatility (which are, in turn, responsible for the large returns). For the 
$q_{n}-$Gaussian noise case, although some discrepancy exits for small $\sigma ^{2}$, the tail is remarkably good.

\section{Degree of dependence between successive returns}

\label{kl}As stated in Sec.~\ref{garch-model}, stochastic variables, $\left\{z_{t}\right\} $, 
in a $GARCH$ process are uncorrelated (see Eq.~(\ref{z})). However, if we combine Eqs.~(\ref{z}) 
and (\ref{garch}), we can verify
that they are not independent. More specifically, for $GARCH( 1,1) $ ($\bar{\sigma}^{2}=1$), we have
\begin{equation}
z_{t}=\sqrt{1+b\left( z_{t-1}^{2}-1\right) +c\left( \frac{z_{t-1}^{2}}{%
\omega _{t-1}^{2}}-1\,\right) }\,\omega _{t} \,.  \label{ztzt1}
\end{equation}
One of the possible measures of the dependence between the $(
z_{t};z_{t-1}) $ stochastic variables consists in using a generalised form of the 
Kullback-Leibler relative entropy \cite{tsallis-kl}, namely 
\begin{eqnarray}
%\begin{array}{c}
I_{q^{\prime }}( p_{1},p_{2})  &\equiv&   
%\frac{1}{1-q^\prime}   \int 
%(p_{1}( u))^{q^\prime}\left[ p_1( u) ^{1-q^{\prime }}-p_2(u) ^{1-q^{\prime }}\right] \,du   \nonumber \\
%&=&
-\int p_{1}( u) \,\ln _{q^{\prime }}\left[ \frac{p_2(u) }{p_1( u) }\right] \,du \,,
%\end{array}
\label{iq2d}
\end{eqnarray}
where $\ln _{q^{\prime }}\left( x\right) \equiv \frac{x^{1-q^{\prime }}-1}{%
1-q^{\prime }}$ ($q^{\prime }$-logarithm). In the limit $q^{\prime
}\rightarrow 1$, $I_{q^{\prime }}$ recovers the standard Kullback-Leibler form \cite{kullback-leibler}. This generalised relative entropy equals zero
whenever $p_2( u) =p_1( u) $, and has the same sign
as $q^\prime $ otherwise. With these properties we can use $I_{q^{\prime
}}$ (with $q^{\prime }>0$) as a way to compute the ``distance", in probability space, from $
p_2( u) $ to $p_1( u) $.
Assume that $u$ is a two dimensional random variable $u\equiv \left(
x,y\right) $ so that $p_{1}( x,y) $ represents the {\it joint
distribution} of $(x,y)$, and $p_2( x,y)  \equiv h_1(x) \,h_2( y) $, where  
 $h_{1}( x)  \equiv \int p_{1}(
x,y) \,dy$ and $h_2( y)  \equiv \int p_1( x,y) \,dx$ are the {\it marginal distributions}. Random variables $x$ and $y$ are independent if $p_{1}( x,y)
=p_{2}( x,y) $. With the functional $I_{q^{\prime}}$ we can measure the
degree of dependence {\it via} distance between probabilities $
p_{1}( x,y) $ and $p_{2}( x,y) $. 
\begin{figure}[tbp]
\includegraphics[width=0.9\columnwidth,angle=0]{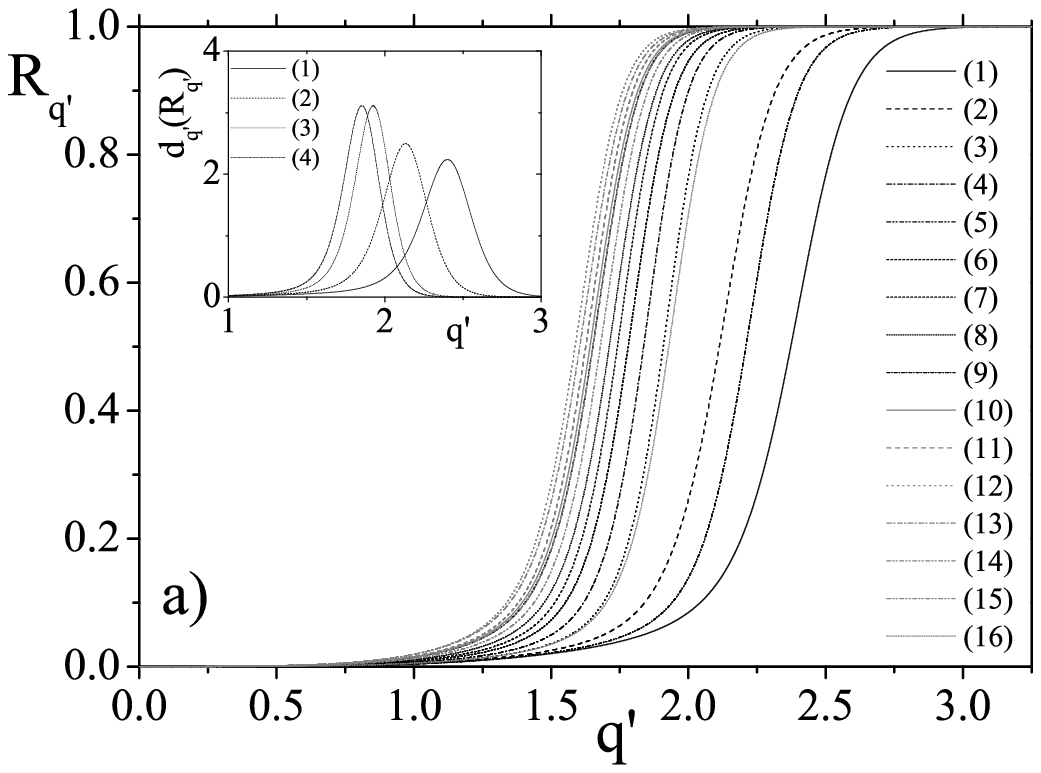}
\includegraphics[width=0.95\columnwidth,angle=0]{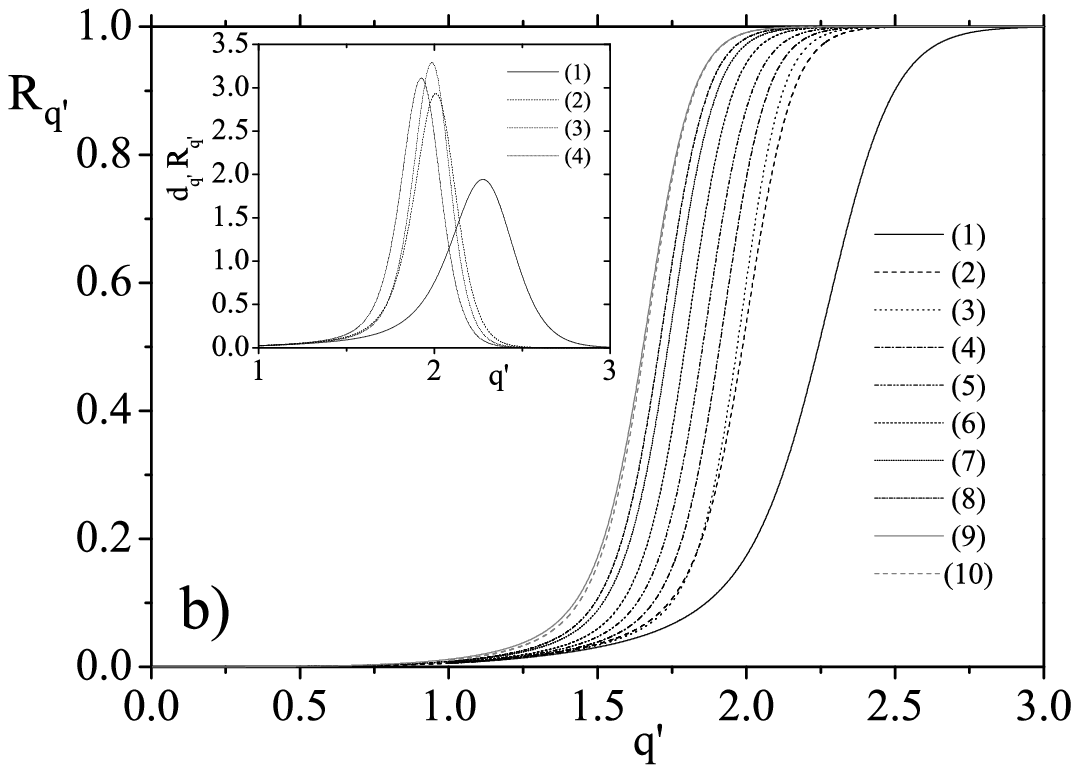}
\caption{Dependence criterion $R_{q^{\prime }}$ vs. $q^{\prime }$ for
various $GARCH(1,1)$ process with typical $\left( q_{n},b,c\right) $ triplets. 
In (a) $q_{n}=1$ and $(b,c)$ have been chosen as follows: 1-$\left( 0,0\right) $, 
2-$( 0.05,$$0) $,3-$( 0.1,$$0.2) $,4-$( 0.15,$$0) $,5-$( 0.2,$$0) $,6-$( 0.2,$$0.2) $,7-$(0.25,$$0) $, 
8-$\left(0.3,0\right) $, 
9-$\left( 0.4,0\right) $, 
10-$\left( 0.4,0.1\right) $, 
11-$\left( 0.4,0.2\right) $,  
12-$\left( 0.4,0.4\right) $, 
13-$\left(0.5,0\right) $, 
14-$\left( 0.2,0.688\right) $, 
15-$\left( 0.35,0\right) $, 
16-$\left( 0.1,0\right) $. 
In (b) $q_{n}=1.2$ and $(b,c)$ have been chosen as follows: 1-$\left( 0,0\right) $, 2-$\left( 0.1,0\right) $, 
3-$\left( 0.1,0.1\right) $, 4-$\left( 0.15,0\right) $, 
5-$\left( 0.2,0\right) $, 6-$\left( 0.25,0\right) $, 7-$\left( 0.3,0.1\right) $, 
8-$\left(0.377,0\right) $, 9-$\left( 0.3,0.45\right) $, 10-$\left( 0.48,0.0\right) $. 
The insets contain the derivative  $dR_{q^{\prime }}/dq^{\prime }$(numerically obtained)
for the first four curves as mere illustration.}
\label{fig-9}
\end{figure}
For this case, $u\equiv \left( x,y\right) $ and $q^{\prime }>0$, it was shown \cite{borland-plastino-tsallis} 
that, besides a lower bound $I_{q^{\prime }}=0$
(total independence of $x$ and $y$), $I_{q^{\prime }}$ presents, for every
value of $q^{^{\prime }}$, an upper bound (complete dependence between $x$
and $y$) given by,
\begin{eqnarray}
%\begin{array}{c}
I_{q^\prime}^{MAX}( p_{1},p_{2}) =
- \int \int [p_{1}( x,y)]^{q^\prime} 
\{ \ln _{q^\prime}h_{1}( x)  \nonumber  \\ 
 +( 1-q^\prime) [\ln _{q^{\prime }}h_{1}( x)][ \ln _{q^{\prime }}h_{2}(y)] \}  \,dx\,dy \,.
%\end{array}
\label{iqmax}
\end{eqnarray}
Dividing Eq.~(\ref{iq2d}) by (\ref{iqmax}) we define
\begin{equation}
R_{q^\prime} \equiv \frac{I_{q^{\prime }}}{I_{q^{\prime }}^{MAX}}\quad \in  \left[ 0,1\right] \,.
\label{rq}
\end{equation}
This ratio can be used as a criterion for measuring the degree of dependence
between random variables. Indeed, it presents an optimal $q^{\prime }$, noted $q^{op}$ \cite{nota1}, for which the 
sensitivity of $R_{q^\prime}$ is maximal ($q^{op}$ corresponds to the inflexion point of  
$R_{q^{\prime }}(q^{\prime })$). Higher (lower) values of  $q^{op}$ represents lower (higher) 
degree of dependence \cite{borland-plastino-tsallis}. 
Taking $x=z_{t}$ and $y=z_{t-1}$ generated from Eq.~(\ref{ztzt1}), and
applying Eq.~(\ref{rq}), we obtained the curves presented in Fig.~\ref{fig-9}
for typical values of $(b,c,q_n)$. For each set  we determined $
q^{op}$ and plotted it {\it versus} $q$ obtained from Eq.~(\ref{ansatz}): see
Fig.~\ref{fig-10}. For both noises that have been illustrated, $q^{op}$ monotonically decreases with $q$. For fixed $q_{n}$, this 
(decreasing) curve does not
depend on the values of $(b,c)$. An illustration of this independence is indicated in Fig.~\ref{fig-10} by using different pairs $(b,c)$ 
that give the same $q$. 
This fact suggests the existence of a  relation between non-Gaussianity 
(represented by $q$), the degree of dependence quantified with $q^{op}$, and the noise index $q_{n}$. 
This triangular relation ($q$,$q^{op}$,$q_{n}$) is analogous to another one which could relate the sensitivity, relaxation and stationarity in weakly chaotic  
systems such as those in which long-range interactions are assumed \cite{tsallis-villasimius}. 
\begin{figure}[tbp]
\includegraphics[width=0.75\columnwidth,angle=0]{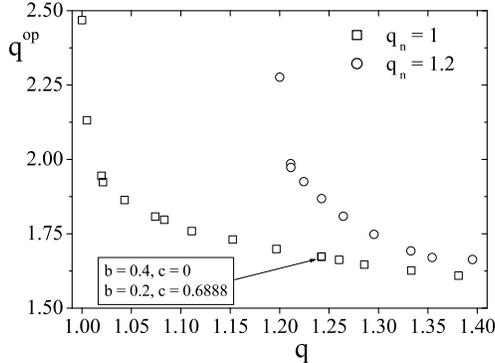}
\caption{Plot of $q^{op}$ vs. $q$ for typical $( q_n,b,c) 
$ triplets. The arrow points two examples which were obtained from
{\it different} triplets, and nevertheless coincide in what concerns the resulting point $(q,q^{op})$.}
\label{fig-10}
\end{figure}

\section{Concluding remarks}

\label{remarks}In this article, we presented a study of the stationary statistical features of the 
$GARCH(1,1)$ stochastic process, which is characterised by the fact that it exhibits a time dependent (and correlated)  
volatility. This attribute makes this system similar to those exhibiting fluctuations in some 
intensive parameter (e.g. temperature), which is at the basis of superstatistics. Inspired by the intimate connection 
between superstatistics and nonextensive statistical mechanics, we obtained an expression linking 
the dynamical parameters $(b,c)$ and the noise nature, $q_{n}$,  with the entropic index $q$ characterising the stationary distribution 
for the associated $GARCH(1,1)$ process. After numerically testing the validity of the approach for the return distributions, we derived analytical expressions 
for the squared volatility stationary distribution in the presence of $q_n$-Gaussian noise. The results are satisfactory for
$q_{n}\simeq 1$. Then, using the  $q$-generalised Kullback-Leibler relative entropy, we  
quantified the degree of dependence between successive returns. This analysis led to an entropic 
index $q^{op}$ which is optimal in the sense that the ratio (33) exhibits maximal sensitivity. We then verified the existence of a direct relation between 
$q^{op}$, the non-Gaussianity, $q$, and the nature of the noise, $q_{n}$.  An interesting property emerged, namely that, whatever be the pair $(b,c)$ that results in a certain $q$ for the stationary distribution, one obtains the same value of $q^{op}$. Consequently, the 
time series will present the same degree of dependence. 
The connection between various entropic indices reminds us of the dynamical scenario within which nonextensive statistical mechanics is formulated. Indeed, various entropic indices emerge therein, which coalesce onto the same value $q=1$ if ergodicity is satisfied.
In the present context, this connection can be analysed as follows. Due to the dynamical characteristics of $GARCH(1,1)$, non-Gaussian distribution for returns ($q \neq 1$) comes from temporal dependence on its second-order moment which is a self-correlated variable. This 
correlation, sign of memory in the process, is the responsible for the breakdown of independence  \cite{tsallis-erice} between  
$z_{t}$ and $z_{t+1}$,  which in turn reflects on the entropic index 
$q^{op}$. The connection between $q$ and $q^{op}$ opens the door to the establishment of a relation between the 
topologies of phase space and probability space. A careful analysis 
of other kind of systems (e.g. long-range hamiltonian models, Langevin-like dynamics obeying generalised 
correlated Fokker-Plank equations) should give a deep insight onto this question.      

\bigskip
L.G. Moyano is acknowledged for fruitful comments on the manuscript.
Partial financial support from Faperj, CNPq, PRONEX/MCT (Brazilian agencies) and 
FCT/MES (Portuguese agency; contract SFRH/BD/6127/2001) is acknowledged as well.

\begin{table}
\begin{center}
\begin{tabular}{|c|c|c|c|c|c|}
\hline
$q_{n}$ & $b$ & $c$ & $\left\langle z^{6} \right\rangle _{numerical}$ & $\left\langle z^{6} \right\rangle 
_{ansatz}$ & error ($\%$) \\ 
\hline\hline
1 & 0.1 & 0.2 & 13.97 & 14.05 & 0.60 \\ \hline
1 & 0.1 & 0.88 & 66.75 & 65.46 & 1.93 \\ \hline
1 & 0.4 & 0.1 & 61.45 & 62.33 & 1.44 \\ \hline
1 & 0.4 & 0.4 & 591.91 & 517.67 & 1.75 \\ \hline
1.2 & 0.1 & 0.1 & 49.41 & 49.08 & 0.67 \\ \hline
1.2 & 0.1 & 0.5 & 55.93 & 55.43 & 0.89 \\ \hline
1.2 & 0.3 & 0.25 & 181.48 & 179.44 & 1.13 \\ \hline
1.2 & 0.3 & 0.45 & 1416.41 & 1455.37 & 2.75 \\ \hline
\end{tabular}
\end{center}
\caption{Percentual error in the {\it sixth}-order moment between numerical and ansatz PDFs 
presented in Figs.~\ref{fig-5} and \ref{fig-6}.
}
\label{tab-1}
\end{table}

\end{multicols}


\begin{references}
\bibitem{transfvar} Since $\sigma _{t}$ and $\omega _{t}$ are uncorrelated and independent, for a given value 
$\sigma $, we can do a change of variable,  Eq.~(\ref{z}), which transforms Eq.~(\ref{pnoise}) into the conditional probability
Eq.~(\ref{pzq-qn}).
\bibitem{nota1}  Optimal in the sense that its two extreme values correspond
to full independence/dependence \cite{borland-plastino-tsallis}.
\end{references}
\end{document}